# Naturalistic Tasks Can Benefit Psychiatry

*Commentary on "Beyond simple laboratory studies: Developing sophisticated models to study rich behavior" by Maselli et al. 2023*

Assia Chericoni and Benjamin Yost Hayden

**Affiliations**
    Department of Neurosurgery
    Baylor College of Medicine
    Houston, TX, USA
    77401

**Lead Contact:**
    Assia Chericoni
    Department of Neurosurgery, Program in Neuroscience
    Baylor College of Medicine
    assia.chericoni@bcm.edu

# MAIN TEXT

Maselli et al. (2023) do an excellent job highlighting how studying rich, complex, naturalistic behaviors can help neuroscientists and psychologists understand neural and mental phenomena. We fully concur with their arguments, as well as with somewhat similar ones made by other scholars (e.g., Gomez-Marin and Ghazanfar, 2019; Krakauer et al., 2017; Niv, 2021; Pearson et al., 2014; Yoo et al., 2021). Indeed, we believe that the potential value of naturalistic paradigms is even greater than the authors state. We are particularly sanguine about the power of the naturalistic approach to help with understanding psychiatric diseases.

In recent years, psychiatrists have become increasingly interested in using tasks developed by cognitive psychologists to quantify dimensions of cognition, such as cognitive control, working memory span, or attentional abilities (e.g., Huys et al., 2016; Montague et al., 2012; Paulus et al., 2016). These dimensions of cognition are hypothesized to correspond to faculties that are specifically impaired in psychiatric disease. Using standard laboratory tasks, then, is useful for improving psychiatric diagnosis and for guiding selection of treatment. Indeed, this *constituent element* viewpoint is the guiding philosophy of multiple modern classification approaches, as embodied in both the DSM-V and the RDoC approach (American Psychiatric Association, 2013; Insel et al., 2010). Critical to the use of these approaches is the use of simple standardized laboratory tasks, such as the Stroop Task, the N-back task, and the digit span task. However, we believe that this approach is intrinsically limited, for several reasons, and that these limitations can be circumvented by using more naturalistic tasks.

To give one example of the difference between a laboratory task approach and a more naturalistic approach, consider cognitive control (Matsumoto and Tanaka, 2004; Miller and Cohen, 2001; Shenhav et al., 2017). Many psychiatric diseases, including depression, addiction, and anxiety disorders, are characterized by impaired cognitive control. Psychiatrists have begun to make use of standardized cognitive control tasks, such as the Stroop Task and the Flanker task to quantify individuals' ability to exert cognitive control (e.g., Epp et al., 2012; Joyal et al., 2019; Stasak et al., 2021). An example of a more naturalistic cognitive control task would be a prey-pursuit task: an agent moves a joystick to move an avatar around on a screen (like Pac-Man) to avoid predators and try to capture prey (Yoo et al., 2020; Yoo et al., 2021). This task requires control - the player must continuously monitor the positions of the self, prey, and predator, and adjust behavior to reach a goal. Unlike the simpler tasks, this control is continuous - it unfolds over time - and it is more naturalistic - prey-pursuit is a classic primate foraging behavior (Sussman et al., 2013).

We see several benefits to using naturalistic tasks like the prey-pursuit task. Perhaps the strongest reason is their greater *dimensionality*. Standard cognitive psychology tasks can provide only low-dimensional measures, and often only a single dimension. However, cognition is likely multifarious; its many dimensions almost certainly include axes that do not readily correspond to the features used in psychology tasks. This is likely to be true in general, and even within the narrow domain of control (Badre et al., 2021; MacDowell et al., 2022). Thus, focusing on standard laboratory tasks represents a form of searching under the streetlight because that's where the light is. Because we don't have a good understanding of the structure of cognition, especially as it relates to disease, it is best to use tasks that explore as many dimensions as possible to try to understand the system.

Besides higher dimensionality, naturalism is itself likely to be valuable. Behavior is likely to be more externally valid in naturalistic tasks (Kingstone et al., 2008; Osborne-Crowley, 2020; Pearson et al., 2014; Sonkusare et al., 2019). For example, it may be that cognitive control is deployed in unnatural ways in the Stroop task, which is highly artificial, but in a task that is based on our natural behavioral repertoires, we are more likely to get at behavioral computations as they are evolved to work (Cisek and Hayden, 2022; Krakauer et al., 2017; Pearson et al., 2014). Moreover, the lack of external validity in simple laboratory tasks means they require a second stage of investigation aimed at justifying their validity. Conversely, naturalistic tasks do not (Burgess et al., 2006).

Likewise, naturalistic stimuli, which probe cognition in a more natural way, might elicit the same compensatory strategies that the patients can adopt during everyday life (i.e., avoidance of specific stimuli that can trigger an obsession, or engaging safety actions to reduce anxiety or distress). Their measurement would be beneficial for both diagnosis and treatment, for instance in patients with ADHD or autism (Canela et al., 2017; Livingston et al., 2019). Of course, this factor is a double-edged sword - it means that researchers must be actively on the lookout for these compensatory strategies - however, because such strategies are likely part of our natural repertoires, including them in the scientific method will likely ultimately be beneficial.

More speculatively, naturalistic tasks may also help circumvent an important problem with simpler tasks known as the reliability paradox (Hedge et al., 2018; Zorowitz and Niv, 2023). That is, tasks that have high reliability tend to have low across-subjects variability. In other words, with conventional tasks, you can often get reliability (within subjects) or variability (across subjects), but not both. But we need both: tasks need to be reliable to be clinically valid, but need to be variable to provide differential diagnosis. There are several possible causes of the reliability paradox; however the simplicity of the task designs necessary to produce reliable results across-subjects is likely to be a major culprit (Hedge et al., 2018). Therefore we would hope that naturalistic tasks may offer an escape path from this paradox.

Another benefit of naturalistic stimuli is that they can elicit multimodal integration and thereby increase the signal-to-noise ratio of brain data (Honey et al., 2012; Schultz and Pilz, 2009; Sonkusare et al., 2019). Thus, inducing the activation of multiple brain areas would make more likely to catch abnormalities and would encompass more of the spectrum of potential symptoms without the need for various distinct tasks.

None of this is to say that naturalistic tasks can solve every problem. Indeed, the tasks have their drawbacks. These include, for example, the fact that they require complex and/or difficult analysis, they require more careful thought in their design, they require new methods to deal with the larger number of correlated nuisance variables, and they require a more sophisticated ethological perspective to interpret. Moreover, they often require more sophisticated sensors - while a simple button press may suffice for a Stroop task, a more naturalistic task may require tracking of full body position (Anderson and Perona, 2014; Bialek, 2022). Fortunately, the hardware and software necessary for such purposes are becoming more readily available (e.g., Bala et al., 2020; Berman et al., 2014; Lin et al., 2014; Marshall et al., 2021; Mathis et al., 2018; Pereira et al., 2019). All together, these are not trivial limitations. However, in our view, the potential benefits to psychiatry are worth paying these costs.